\documentclass[aps,prl,reprint,showpacs,floatfix,superscriptaddress]{revtex4-1}

\usepackage{graphicx}  

\usepackage{amsmath}

\newcommand{\fpi}{\mbox{$F_\pi$}}

\begin{document}

\title{Unique Access to $u$-Channel Physics: Exclusive Backward-Angle Omega Meson Electroproduction}
 

\author{W.B.~Li}
\affiliation{University of Regina, Regina, Saskatchewan S4S 0A2, Canada}
\affiliation{College of William and Mary, Williamsburg, Virginia 23185}
\author{G.M.~Huber}
\affiliation{University of Regina, Regina, Saskatchewan S4S 0A2, Canada}
\author{H.P.~Blok}
\affiliation{VU University, NL-1081 HV Amsterdam, The Netherlands}
\affiliation{NIKHEF, Postbus 41882, NL-1009 DB Amsterdam, The Netherlands}
\author{D.~Gaskell}
\affiliation{Thomas Jefferson National Accelerator Facility, Newport News, Virginia 23606}
\author{T.~Horn}
\affiliation{Catholic University of America, Washington, DC 20064}
\author{K.~Semenov-Tian-Shansky}
\affiliation{National Research Centre Kurchatov Institute: Petersburg Nuclear Physics Institute, RU-188300 Gatchina, Russia}
\affiliation{Saint Petersburg National Research Academic University of the Russian Academy of Sciences, RU-194021 St. Petersburg, Russia}
\author{B.~Pire}
\affiliation{CPHT, CNRS, \'Ecole Polytechnique, IP Paris, 91128-Palaiseau, France}
\author{L.~Szymanowski}
\affiliation{National Centre for Nuclear Research (NCBJ),  02-093 Warsaw, Poland}
\author{J.-M.~Laget}
\affiliation{Thomas Jefferson National Accelerator Facility, Newport News, Virginia 23606}
\author{K.~Aniol}
\affiliation{California State University Los Angeles, Los Angeles, California 90032}
\author{J.~Arrington}
\affiliation{Physics Division, Argonne National Laboratory, Argonne, Illinois 60439}
\author{E.J.~Beise}
\affiliation{University of Maryland, College Park, Maryland 20742}
\author{W.~Boeglin}
\affiliation{Florida International University, Miami, Florida 33119}
\author{E.J.~Brash}
\affiliation{Christopher Newport University, Newport News, Virginia 23606}
\author{H.~Breuer}
\affiliation{University of Maryland, College Park, Maryland 20742}
\author{C.C.~Chang}
\affiliation{University of Maryland, College Park, Maryland 20742}
\author{M.E.~Christy}
\affiliation{Hampton University, Hampton, Virginia 23668}
\author{R.~Ent}
\affiliation{Thomas Jefferson National Accelerator Facility, Newport News, Virginia 23606}
\author{E.F.~Gibson}
\affiliation{California State University, Sacramento, California 95819}
\author{R.J.~Holt}
\affiliation{Caltech, Pasadena, California 91125}
\author{S.~Jin}
\affiliation{Kyungpook National University, Daegu, 702-701, Republic of Korea}
\author{M.K.~Jones}
\affiliation{Thomas Jefferson National Accelerator Facility, Newport News, Virginia 23606}
\author{C.E.~Keppel}
\affiliation{Hampton University, Hampton, Virginia 23668}
\affiliation{Thomas Jefferson National Accelerator Facility, Newport News, Virginia 23606}
\author{W.~Kim}
\affiliation{Kyungpook National University, Daegu, 702-701, Republic of Korea}
\author{P.M.~King}
\affiliation{Ohio University, Athens, OH 45701 }
\author{V.~Kovaltchouk}
\affiliation{Ontario Tech University, Oshawa, Ontario L1G 0C5, Canada}
\author{J.~Liu}
\affiliation{Shanghai Jiao Tong University, Shanghai 200240, China}
\author{G.J.~Lolos}
\affiliation{University of Regina, Regina, Saskatchewan S4S 0A2, Canada}
\author{D.J.~Mack}
\affiliation{Thomas Jefferson National Accelerator Facility, Newport News, Virginia 23606}
\author{D.J.~Margaziotis}
\affiliation{California State University Los Angeles, Los Angeles, California 90032}
\author{P.~Markowitz}
\affiliation{Florida International University, Miami, Florida 33119}
\author{A.~Matsumura}
\affiliation{Tohoku University, Sendai, Japan}
\author{D.~Meekins}
\affiliation{Thomas Jefferson National Accelerator Facility, Newport News, Virginia 23606}
\author{T.~Miyoshi}
\affiliation{Tohoku University, Sendai, Japan}
\author{H.~Mkrtchyan}
\affiliation{A.I. Alikhanyan National Science Laboratory, Yerevan 0036, Armenia}
\author{I.~Niculescu}
\affiliation{James Madison University, Harrisonburg, Virginia 22807}
\author{Y.~Okayasu}
\affiliation{Tohoku University, Sendai, Japan}
\author{L.~Pentchev}
\affiliation{Thomas Jefferson National Accelerator Facility, Newport News, Virginia 23606}
\author{C.~Perdrisat}
\affiliation{College of William and Mary, Williamsburg, Virginia 23187}
\author{D.~Potterveld}
\affiliation{Physics Division, Argonne National Laboratory, Argonne, Illinois 60439}
\author{V.~Punjabi}
\affiliation{Norfolk State University, Norfolk, Virginia 23504}
\author{P.E.~Reimer}
\affiliation{Physics Division, Argonne National Laboratory, Argonne, Illinois 60439}
\author{J.~Reinhold}
\affiliation{Florida International University, Miami, Florida 33119}
\author{J.~Roche}
\affiliation{Ohio University, Athens, OH 45701 }
\author{P.G.~Roos}
\affiliation{University of Maryland, College Park, Maryland 20742}
\author{A.~Sarty}
\affiliation{Saint Mary's University, Halifax, Nova Scotia B3H 3C3 Canada}
\author{G.R.~Smith}
\affiliation{Thomas Jefferson National Accelerator Facility, Newport News, Virginia 23606}
\author{V.~Tadevosyan}
\affiliation{A.I. Alikhanyan National Science Laboratory, Yerevan 0036, Armenia}
\author{L.G.~Tang}
\affiliation{Hampton University, Hampton, Virginia 23668}
\affiliation{Thomas Jefferson National Accelerator Facility, Newport News, Virginia 23606}
\author{V.~Tvaskis}
\affiliation{VU university, NL-1081 HV Amsterdam, The Netherlands}
\affiliation{NIKHEF, Postbus 41882, NL-1009 DB Amsterdam, The Netherlands}
\author{J.~Volmer}
\affiliation{VU university, NL-1081 HV Amsterdam, The Netherlands}
\affiliation{DESY, Hamburg, Germany}
\author{W.~Vulcan}
\affiliation{Thomas Jefferson National Accelerator Facility, Newport News, Virginia 23606}
\author{G.~Warren}
\affiliation{Pacific Northwest National Laboratory, Richland, Washington 99352}
\author{S.A.~Wood}
\affiliation{Thomas Jefferson National Accelerator Facility, Newport News, Virginia 23606}
\author{C.~Xu}
\affiliation{University of Regina, Regina, Saskatchewan S4S 0A2, Canada}
\author{X.~Zheng}
\affiliation{University of Virginia, Charlottesville, Virginia 22904}
\collaboration{The Jefferson Lab \fpi\ Collaboration}
\noaffiliation

\date{\today}



\begin{abstract}


Backward-angle meson electroproduction above the resonance region, which was
previously ignored, is anticipated to offer unique access to the three quark
plus sea component of the nucleon wave function. In this letter, we present the
first complete separation of the four electromagnetic structure functions above
the resonance region in exclusive $\omega$ electroproduction off the proton,
$ep\rightarrow e^{\prime}p\omega$, at central $Q^2$ values of 1.60, 2.45
GeV$^2$, at $W=2.21$ GeV.  The results of our pioneering $-u\approx -u_{min}$
study demonstrate the existence of a unanticipated backward-angle cross section
peak and the feasibility of full L/T/LT/TT separations in this never explored
kinematic territory. At $Q^2$=2.45 GeV$^2$, the observed dominance of
$\sigma_T$ over $\sigma_L$, is qualitatively consistent with the collinear QCD
description in the near-backward regime, in which the scattering amplitude
factorizes into a hard subprocess amplitude and baryon to meson transition
distribution amplitudes (TDAs): universal non-perturbative objects only
accessible through backward angle kinematics.

\end{abstract}

\maketitle

\begin{figure}
\begin{center}
\includegraphics[width=.49\linewidth]{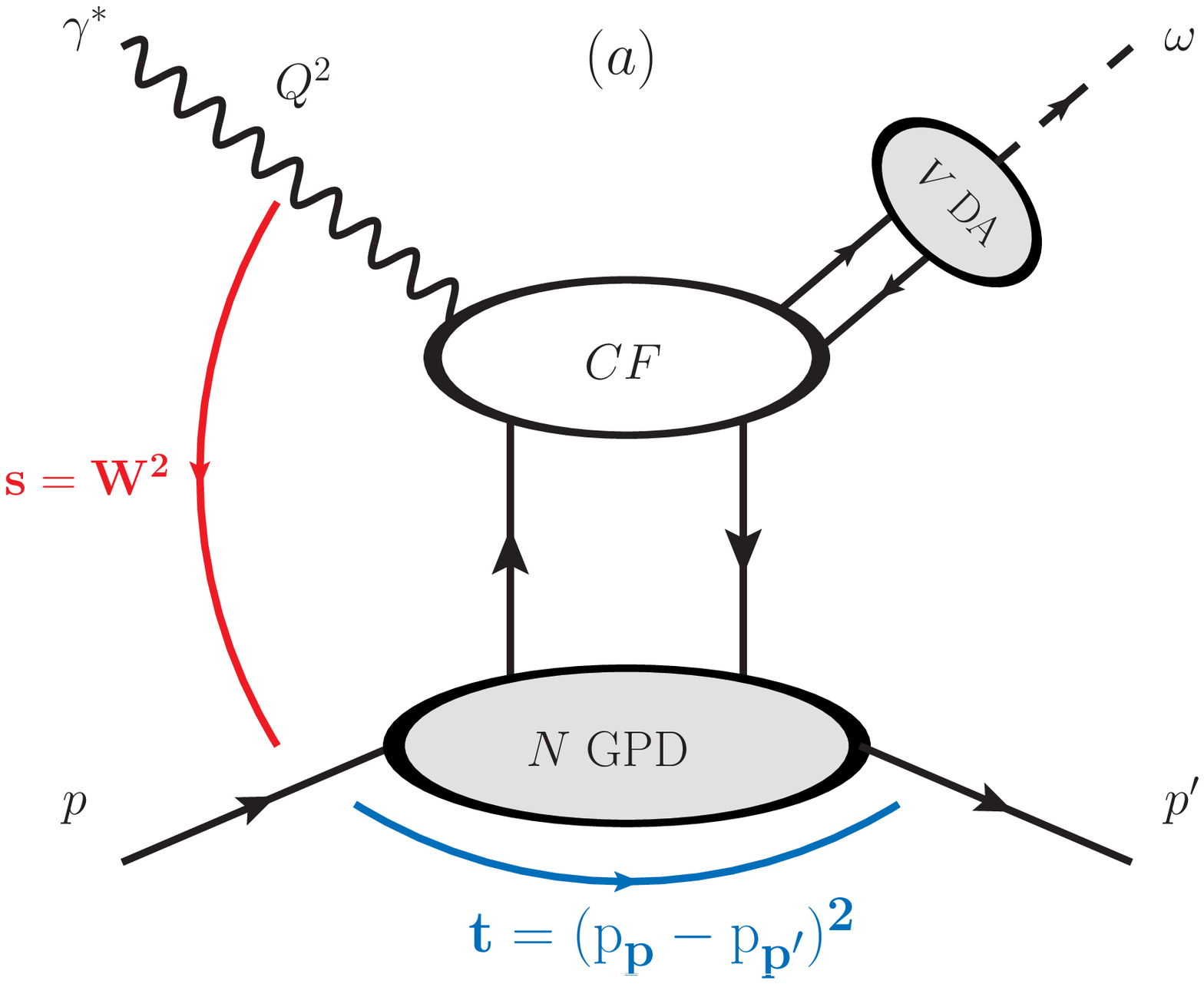}~
\includegraphics[width=.49\linewidth]{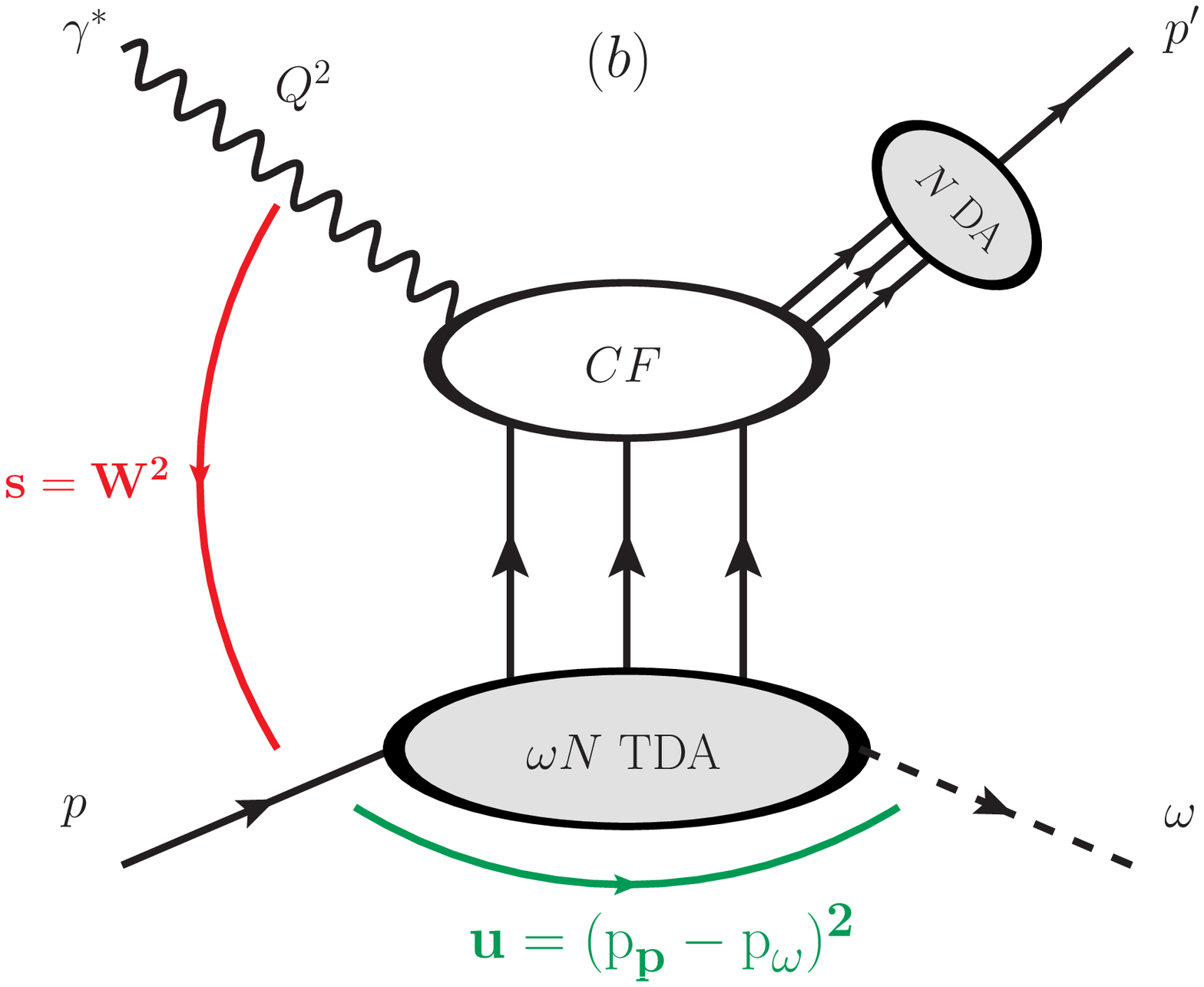}
\end{center}
%
\caption{QCD collinear factorization mechanisms for exclusive $\omega$ electroproduction off a proton (p) probed by $\gamma^*$ at large $Q^2$ and $W$ : (a) Forward regime (small $-t$), GPDs (bottom oval) and the $\omega$-DA(top-right oval); (b) Backward regime (small $-u$), $\omega N$ TDAs (bottom oval) and the proton $N$-DA(top-right oval).}
\label{fig:GPD_TDA}
\end{figure}

Deep exclusive reactions have recently gained much attention, as they provide direct access to the internal structure of hadrons.  Measurements of such reactions at different squared four-momenta of the exchanged virtual photon ($\gamma^*$), $Q^2$, and at different hadron four-momentum transfer, Mandelstam variable $t$ and $u$ (defined in Fig.~\ref{fig:GPD_TDA}), are used to probe QCD's transition from hadronic degrees of freedom at the long distance scale to quark-gluon degrees of freedom at the short distance scale.


The standard experimental configuration to probe deep exclusive reactions involves accelerated charged lepton collisions with a hydrogen target.  While most experiments detect the scattered leptons and forward going final state particles (in the laboratory reference frame), the reaction of interest of this letter concerns final state particles produced at backward angle. The visualization of the backward-angle interaction gives rise to a unique physical picture: a target proton absorbs most of the momentum transfer (by $\gamma^*$), and recoils forward; whereas the produced meson remains close to the target nearly at rest. This type of reaction is sometimes referred to as a ``knocking a proton out of a proton'' process. The backward-angle exclusive observables accessed by the methodology presented in this letter, opens up new opportunities to extend the current knowledge on the nucleon structure to an unexplored kinematic region.

In the Bjorken limit (sufficiently large $Q^2$ and invariant mass $W$, and $-t/Q^2 \ll 1$), the longitudinal scattering amplitude factorizes into a hard scattering perturbative contribution, and soft Generalized Parton Distributions (GPDs) of the nucleon and distribution amplitudes (DAs) of the meson. The vector meson ($\omega$) production through the GPD in the near-forward kinematics is demonstrated in Fig.~\ref{fig:GPD_TDA} (a). GPDs are light-cone matrix elements of non-local bilinear quark and gluon operators that describe the three-dimensional structure of hadrons, by correlating the internal transverse position of partons to their longitudinal momentum.  For a review, see Refs.~\cite{mueller94, radyushkin96, ji97a, ji97b, goeke01, diehl03, belitsky05, boffi07, guidal13, kumericki16, frankfurt99}.




Analogous universal structure functions accessible in ``near-backward'' kinematics are known as baryon-to-meson Transition Distribution Amplitudes (TDAs)~\cite{Pire:2004ie, Pire:2005ax, lansberg11, pire15, kirill15}, see Fig.~\ref{fig:GPD_TDA} (b), which are light-cone matrix elements of non-local three quark operators. In the TDA picture, the backward-angle meson is produced as the $\gamma^*$ probes the meson cloud structure of the nucleon. 



The TDA collinear factorization regime for hard meson production has two key marking signs in near-backward kinematics which can be tested experimentally~\cite{Pire:2004ie, Pire:2005ax, lansberg11, pire15, kirill15}:
\begin{itemize}
\item The dominance of the transverse polarization of the virtual photon results in the suppression of the longitudinal cross section ($\sigma_{L}$) at large $Q^2$ by at least a factor of $1/Q^2$: $\sigma_{L}/\sigma_{T} < \mu^2/Q^2$ and $\sigma_{T} \gg \sigma_{L}$, where $\mu$ is a typical hadronic scale. 

\item The characteristic $1/Q^8$ scaling behavior of the transverse cross section ($\sigma_{T}$) for fixed Bjorken $x$: $x_{\rm B} = \frac{Q^2}{2\textrm{p}_{\textrm{p}} \textrm{q}}$, where $\textrm{p}_{\textrm{p}}$ and $\textrm{q}$ are the four momenta of the virtual proton and $\gamma^*$, respectively.
\end{itemize}
In a recent publication~\cite{park18}, the CLAS collaboration reported the first measurement of the cross sections for exclusive $\pi^+$ electroproduction off the proton in near-backward kinematics. The result gives promising signs of the predicted $1/Q^8$ scaling of the cross section by TDA, however, the critical evidence for $\sigma_{T}$ dominance remains missing.


In this letter, we present a pioneering study of backward-angle $\omega$ cross sections from exclusive electroproduction: $e p \rightarrow e^{\prime} p \omega$ using the missing-mass reconstruction technique. The extracted cross sections are separated into the transverse (T), longitudinal (L), and LT, TT interference terms. This allows for comparing the individual $\sigma_{L}$ and $\sigma_{T}$ contributions to the TDA calculations, and verifying the predicted $\sigma_{T}$ dominance.

The general form of two-fold virtual-photon differential cross section in terms of the structure functions is given:
\begin{equation}
\begin{split}
2 \pi \frac{d^2 \sigma}{dt ~ d\phi} = \frac{d \sigma_{T}}{dt} + 
\epsilon ~ \frac{d \sigma_{L}}{dt} &+ \sqrt{2\epsilon(1+\epsilon)}~ 
\frac{d\sigma_{LT}}{dt} \cos \phi \\
& + \epsilon ~ \frac{d\sigma_{TT}}{dt} \cos 2\phi \,,
\label{eqn:xsection_LT}
\end{split}
\end{equation} 
where $\epsilon$ is the $\gamma^*$ longitudinal polarization $\epsilon = \left(1 + 2 \frac{|\vec{q}|^2}{Q^2} \tan^2\frac{\theta_e}{2} \right)^{-1}$, $\theta_e$ is the scattered electron polar angle; $\phi$ is the azimuthal angle between the electron scattering plane and the proton target reaction plane. For brevity, differential cross sections such as $d\sigma_{T}/dt$ will be expressed as $\sigma_{T}$.  Separating $\sigma_{L}$ from $\sigma_{T}$, and extracting the interference terms relies on an experimental technique known as Rosenbluth separation. This technique requires two measurements at different $\epsilon$ (dependent upon the beam energy and electron scattering angle), while other Lorentz invariant quantities are kept constant. The interference terms, $\sigma_{LT}$ and $\sigma_{TT}$, dictate the azimuthal modulation for a given opening angle $\theta$ between the proton recoil momentum and the $\gamma^*$ momentum.

The analyzed data were part of experiment E01-004 (F$_{\pi}$-2), which used 2.6-5.2~GeV electron beams on a liquid hydrogen target and the high precision particle spectrometers in Jefferson Lab Hall C~\cite{horn06, blok08}. The data set has two central $Q^2$ values: $Q^2=1.60$ and 2.45~GeV$^2$, at common central $W=2.21$~GeV. The primary objective of the experiment was to detect coincidence $e$-$\pi$ at forward-angle, but backward-angle $\omega$ ($e$-$p$) were fortuitously acquired. 



The recoil protons were detected in the High Momentum Spectrometer (HMS), while the scattered electrons were detected in the Short Orbit Spectrometer (SOS). Both spectrometers include two sets of drift chambers for tracking and scintillator arrays for triggering. A detailed description of the experimental configuration is documented in Refs.~\cite{blok08}.
  
In order to select $e^-$ in the SOS, a gas Cherenkov detector containing Freon-12 at 1 atm was used in combination with a lead-glass calorimeter. The positively charged $\pi^+$ were rejected in the HMS using an aerogel Cherenkov detector with refractive index of 1.03. The rare $e^+$ were rejected using a gas Cherenkov detector filled with C$_4$F$_{10}$ at 0.47 atm. Most remaining contamination of the $e$-$p$ events was rejected by a coincidence time cut of $\pm1$~ns. Background originating from the aluminum target cell and random coincidence events, $<$5\% contribution to the total yield, was subtracted from the charge normalized yield. Proton loss due to multiple scattering inside the HMS was estimated as 7-10\%~\cite{wenliang17}.


\begin{figure}[t]
\centering
\includegraphics[scale=0.43]{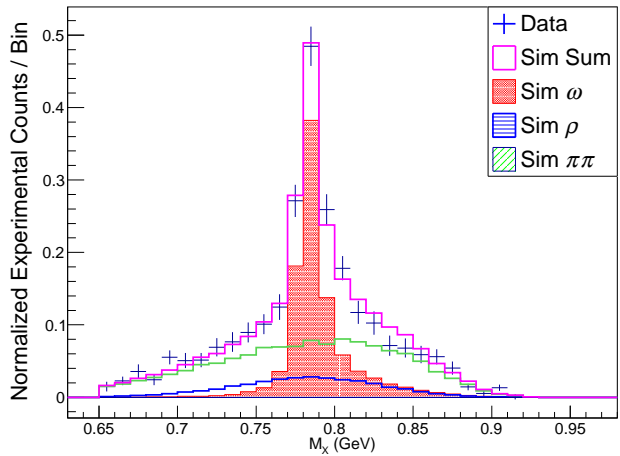}
\caption{Reconstructed missing-mass ($M_X$) for $ep\rightarrow e^{\prime} p X$ at $Q^2 = 2.45$~GeV$^2$ (blue crosses). The simulated distributions for $\rho$ (blue), $\omega$ (red) and $\pi\pi$ (green) are used to describe the measured reaction.}
\label{missmass}
\end{figure}


Unlike the exclusive $\pi^+$ channel \cite{horn06, blok08}, the $\omega$ events sit on a broad background, as shown in the reconstructed missing-mass spectrum for $ep\rightarrow e^{\prime} p X$ in Fig.~\ref{missmass}. The final state particle $X$ could include: $\omega$, $\rho$ or two-$\pi$ production ($\pi\pi$). For each $Q^2$-$\epsilon$-$u$-$\phi$ bin, extracting $\omega$ is a two step process. First, simulations were used to determine the contribution of each final state particle to the $M_X$ distribution. Here, the shape of the distribution for each particle is dictated by the detector acceptance and the particle decay width, while the normalization (scale) factor of the simulated distribution is determined by the fit to the data (simultaneously). In the second step, the background (scaled $\rho$ and $\pi\pi$ simulations) are subtracted from the data to obtain the $\omega$ experimental yield.

Two quality control criteria were introduced to validate the background subtraction procedure: 1. The $\chi^2$ per-degree-of-freedom ($\chi^2/$dof) comparison between the experimental and simulated $\omega$ yields, defined as $Y_{\omega~\textrm{exp}} = Y_\textrm{Data} - Y_{\rho~\textrm{sim}} - Y_{\pi\pi~\textrm{sim}}$, and $Y_{\omega~\textrm{sim}}$; 2. $\chi^2/$dof comparison between the experimental and simulated background yields, defined as $Y_{\textrm{BG}~\textrm{exp}} = Y_\textrm{Data} - Y_{\omega~\textrm{sim}}$ and $Y_{\textrm{BG}~\textrm{sim}}=Y_{\rho~\textrm{sim}} + Y_{\pi\pi~\textrm{sim}}$. Both $\chi^2/$dof distributions obey Poisson statistics with center values: 0.94, 1.3, and widths: 0.77, 0.97, respectively. The detailed analysis procedure is documented in Ref.~\cite{wenliang17}.

\begin{figure}[t]
\includegraphics[width=1\linewidth]{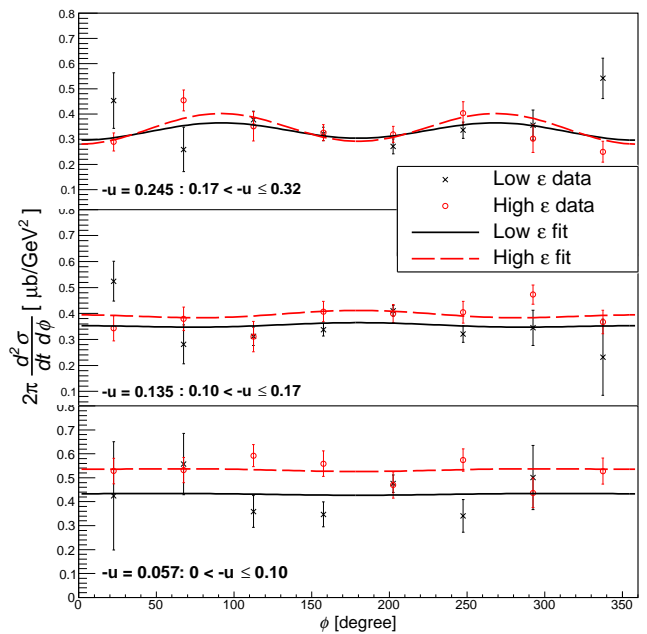}
\caption{Unseparated cross section as function of $\phi$ at $-u=0.057$, 0.135 and 0.245~GeV$^2$ (from bottom to top) at $Q^2=1.6$~GeV$^2$. The higher $\epsilon=0.59$ and lower $\epsilon=0.32$ data are shown in red circles and black crosses, respectively. Red dashed (higher $\epsilon$) and black solid (lower $\epsilon$) lines are the fitting results used in Eq.~\ref{eqn:xsection_LT}. Note that the fitting performed takes into account data at both $\epsilon$ settings simultaneously.}
\label{fig:money}
\end{figure}

For each $Q^2$ setting, two data sets with different $\epsilon$ values were acquired: $Q^2=1.6$ GeV$^2$, $\epsilon=0.32$, $0.59$; at $Q^2=2.45$ GeV$^2$, $\epsilon = 0.27, 0.55$. Data at each $Q^2$-$\epsilon$ setting were divided into three $u$ bins and eight $\phi$ bins. Fig.~\ref{fig:money} shows the unseparated experimental cross section at $Q^2=1.6$~GeV$^2$ as functions of $\phi$ at three $-u$ bins. The separated cross section is obtained from fitting the data at both $\epsilon$ settings simultaneously using Eq.~\ref{eqn:xsection_LT}.


The experimental acceptance covers a range of $Q^2$, $W$ values, thus the measured experimental yields represent an average over the covered range. As a result, each $-u$ bin has a slightly different average value $\overline{Q^2}$ and $\overline{W}$.  In order to minimize errors resulted from the averaging, the experimental cross sections were calculated by comparing the experimental yields to a Monte-Carlo simulation of the experiment. The Monte-Carlo includes a detailed description of the spectrometer acceptance, multiple scattering, energy loss due to ionization, decay and radiative process.

%
%
%
%
%

The shape of the simulated $\omega$ tail is influenced by radiative effects describing the emission of real or virtual photons, and multiple scattering. For more information on the simulation, see Ref.~\cite{blok08}. The matching between the simulated and the experimental resolutions was verified with $ep$ elastic scattering data and the relatively small observed effect is included in the point-to-point systematic uncertainty. Additionally, to avoid sensitivity to some kinematic regions (at larger $\theta$) with large contributions from the radiative events, bins with simulated $\omega$ tail contribution $>$60\% of the overall distribution are excluded from the analysis (9\% of the bins).

The uncertainty in the separated cross sections includes both statistical and systematic contributions. The statistical contribution consists of the error in determining ``good'' $\omega$ from the background subtraction procedure (fitting error included), the uncertainties in detector performance (efficiencies and tracking) and beam characteristics on a run-by-run basis. A comprehensive study was carried out to obtain the total systematic uncertainties for the separated cross section. It includes three parts: 1. Correlated scale error of the unseparated cross section (2.6\%); 2. Point-to-point variations due to the cross section model dependence in simulation; 3. Effects of the error amplification (by a factor of $1/\Delta \epsilon$) of the $\epsilon$ uncorrelated $u$ correlated systematic error (1.7-2.0\%). The effects of all three parts are added in quadrature as the total systematic error and are reported separately for each $u$ bin.

\begin{figure}[t]
\includegraphics[scale=0.44]{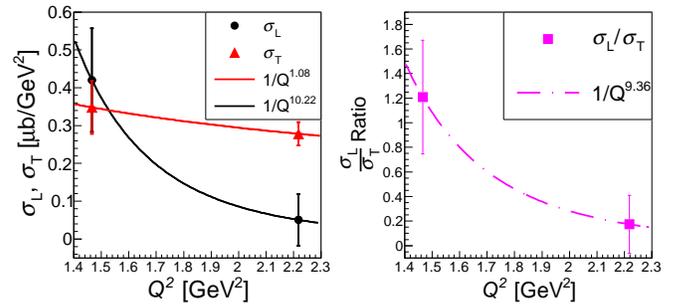}
\caption{Left: $\sigma_{L}(u=u_{\rm min})$ and $\sigma_{T} (u=u_{\rm
    min})$ as function of $Q^2$ for the lowest $-u$ bin. Right: $\sigma_{L}(u=u_{\rm min})/\sigma_{T}(u=u_{\rm
    min})$ ratio as function of $Q^2$. Fitted lines are for visualization
    purpose only.}
\label{fig:ratio}
\end{figure}

To investigate the $Q^2$ dependence, $\sigma_{L}$ and $\sigma_{T}$ for
the smallest $-u$ bin ($u-u_{\rm min}=0$) from the two $Q^2$ settings are
plotted on the left panel of Fig.~\ref{fig:ratio}, whereas the $\sigma_{\rm
  L}/\sigma_{T}$ ratio is plotted on the right. $\sigma_{T}$ shows a
flat $Q^2$ dependence, whereas $\sigma_{L}$ decreases significantly as
$Q^2$ rises. The drop in $\sigma_{L}/\sigma_{T}$ ratio as function of
$Q^2$ is qualitatively consistent with the prediction of TDA collinear
factorization.  

\begin{figure}[t]
\centering
\includegraphics[scale=0.44]{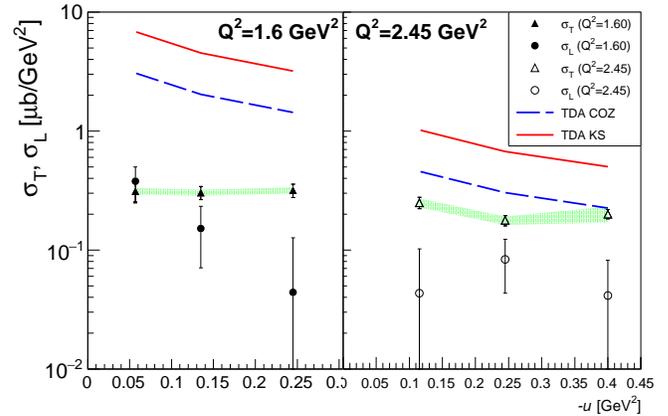}
\caption{$\sigma_{T}$ (triangles), $\sigma_{L}$ (circles) as function
  of $-u$, at $Q^2=1.6$ GeV$^2$ (left), 2.45 GeV$^2$ (right). 
   For the lowest $-u$ bin. 
   TDA predictions for $\sigma_{T}$: COZ~\cite{chernyak89} (blue dashed
   lines) and KS~\cite{king87} (red solid lines). 
   The green bands indicate correlated systematic uncertainties
  for $\sigma_{T}$, the uncertainties for $\sigma_{L}$ have similar
  magnitude.}
\label{fig:sigt}
\end{figure}




The extracted $\sigma_{L}$ and $\sigma_{T}$ as a function of $-u$ at
$Q^2=1.6$ and 2.45~GeV$^2$ are shown in Fig.~\ref{fig:sigt}. 
The two sets of TDA
predictions for $\sigma_{T}$ each assume different nucleon DAs as
input. The predictions were calculated at the specific $\overline{Q^2}$,
$\overline{W}$
values of each $u$ bin. The predictions at three $u$ bins are joined by
straight lines.
At $Q^2 = 2.45$ GeV$^2$, TDA predictions are within the same order of magnitude as the data; whereas at $Q^2 = 1.6$ GeV$^2$, the TDA model over predicts the data by a factor of $\sim$10.  This is very similar to the recent backward-angle $\pi^+$ data from CLAS \cite{park18}, where the TDA prediction is within 50\% of the data at $Q^2$=2.5 GeV$^2$, but far higher than the unseparated data at $Q^2=$1.7 GeV$^2$.  Together, data sets suggest that the boundary where the TDA factorization applies may begin around $Q^2 = 2.5$ GeV$^2$.  





The behavior of $\sigma_{L}$ differs greatly at the two $Q^2$ settings. At $Q^2=$1.6~GeV$^2$, $\sigma_{L}$ falls almost exponentially as a function of $-u$; at $Q^2=$2.45~GeV$^2$, $\sigma_{L}$ is constant near zero (within one standard deviation) and this is consistent with the leading-twist TDA prediction: $\sigma_{L}\approx 0$.





\begin{figure}[ht]
\centering
\includegraphics[scale=0.44]{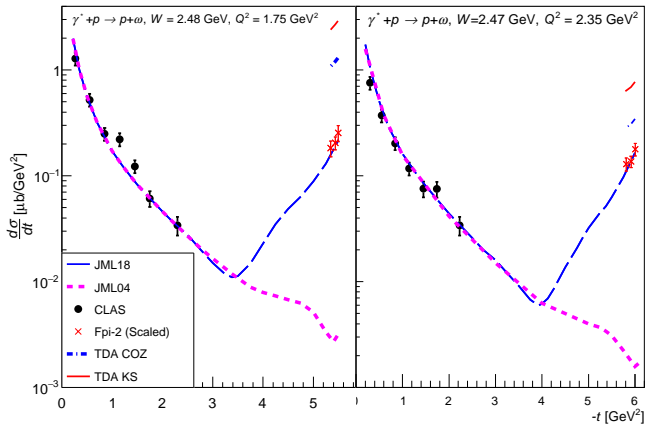}
\caption{Exclusive $\omega$ electroproduction cross section as a function of $-t$
  at $Q^2=1.75$ (left panel) and $Q^2=2.35$~GeV$^2$ (right panel). The CLAS
  data are the black dots in the near-forward kinematics region
  ($-t<2.5$~GeV$^2$), and the F$_\pi$-2 are the red crosses in the backward
  region ($-t>5$~GeV$^2$), scaled to the kinematics of the CLAS data, as
  described in the text. The blue and magenta dashed thick lines are
  Regge trajectory based JML04 and JML18 predictions, respectively. The
  short curves above the F$_\pi$-2 data are TDA predictions based on
  COZ~\cite{chernyak89} (blue solid) and KS~\cite{king87} (red solid) DAs.}
\label{omega_t_slope}
\end{figure}

The combined data from CLAS~\cite{morand05} and F$_{\pi}$-2 cover both forward and
backward-angle kinematics, and jointly form a complete $-t$ evolution picture
for the $e p \rightarrow e^{\prime} p \omega$ reaction. 
The CLAS data, at $W\sim2.48$~GeV$^2$, $Q^2=1.75$ and 2.35 GeV$^2$, are shown
in the left and right panels of Fig.~\ref{omega_t_slope}, respectively.
Because of the similarities in the kinematics, the F$_{\pi}$-2 data (this work)
are scaled to those of the CLAS data.  The $W$
dependence of the backward-angle cross section is unknown, therefore the scaling
procedure: $(W^2-m_p^2)^{-2}$, based on the forward-angle phenomenology studies,
is applied~\cite{brauel79}. 
The $Q^2$ scaling is based on the empirical fit used to extract
the separated cross sections of this work. This empirical model is documented
in Ref.~\cite{wenliang17}. In addition to the scaling, the extracted $-u$
dependent cross section from F$_\pi$-2 is translated to the $-t$ space of the
CLAS data.

Fig.~\ref{omega_t_slope} indicates strong evidence of the existence of the
backward-angle peak at $-t > 5$ GeV$^2$ for both $Q^2$ settings,
with strength $\sim 1/10$ of the forward-angle cross section.
Previously, the ``forward-backward'' peak phenomenon
was only observed in $\pi^+$ photoproduction data~\cite{vgl96,guidal97,anderson69,anderson69b,boyarski68}.
This was successfully interpreted using the Regge 
trajectory based VGL model \cite{vgl96,guidal97}.

The results presented in this paper have demonstrated
that the missing-mass reconstruction technique, in combination with the
high precision spectrometers in coincidence mode at Hall C of Jefferson Lab,
is able to reliably perform a full L/T separation of the backward-angle
exclusive reaction $ep \to e'p \omega$. Since the missing
mass reconstruction method does not require the detection of the produced meson,
this allows the possibility to extend experimental kinematic coverage that was
considered to be inaccessible through the standard direct detection method. If used in
combination with a large acceptance detector, such as CLAS-12, one could
systematically study the complete $t$ evolution of a given interaction,
thus unveiling new aspects of nucleon structure.
The separated cross sections show indications of a regime where
$\sigma_T \gg \sigma_L$
for
$ep \to e'p \omega$,
qualitatively consistent with the TDA factorization approach in backward-angle
kinematics. However, the approach relying on the QCD partonic picture
applying at large enough $Q^2$ involves different mechanisms for the
forward and backward peaks and could not provide a unique description
in the whole range in $-t$.

An alternative description for the $\omega$-meson electroproduction 
cross section is given by the Regge based JML model.
It describes the JLab 
$\pi$ electroproduction cross sections over
a wide kinematic range without destroying good agreement at 
$Q^2=0$~\cite{laget10, laget11}. 
Two  JML model predictions are plotted in Fig. 6: JML04~\cite{laget04} (prior to F$_\pi$-2 data) and JML18. JML04 includes the near-forward Regge contribution at $-t < 1$ GeV$^2$ and $N$-exchange in the $u$-channel with a $t$-dependent cutoff mass. It significantly underpredicts the backward-angle cross section.
%
In JML18~\cite{laget18}, the principle of the $u$-channel treatment is the same as in the $t$-channel neutral pion electroproduction~\cite{laget11}.
It includes, in addition, an estimation of the contribution of the $\rho$-$N$ and $\rho$-$\Delta$ unitarity rescattering (Regge) cuts, allowing an excellent description of the combined data within a unique framework. In particular, the $-u$ dependence and the strength of the backward angle peak are described well at both $Q^2$ settings. The inelastic exchange diagrams are the main sources to the observed backward-angle peak, with one third of the contribution coming from the $\rho^0$-$\omega$ transition, and the rest coming from $\rho^+$-$N$ and $\Delta$ resonance. However, JML18 lacks the prediction of the $Q^2$-dependence of the $\sigma_{L}/\sigma_{T}$ ratio.

In conclusion, the presented experimental data hint on the early onset of the QCD-based factorized description of electroproduction of $\omega$ in the backward kinematics regime for $Q^2$ in the few GeV$^2$ range. This opens a way to the experimental access of nucleon-to-meson TDAs and provides a new window on the quark-gluon structure of nucleons. These data also supply a new interesting testing bench for Regge-based hadronic models.

\begin{acknowledgments}
We acknowledge the excellent efforts provided by the staff of the Accelerator and the Physics Divisions at Jefferson Lab. This work is supported by NSERC (Canada) FRN: SPAPIN-2016-00031, DOE and NSF (USA), FOM (Netherlands), NATO, and NRF (Rep. of Korea).  Additional support from Jefferson Science Associates and the University of Regina is gratefully acknowledged. This material is based upon work supported by the U.S. Department of Energy under contracts DE-AC05-06OR23177 and DE-AC02-06CH11357. L. S. is supported by the grant 2017/26/M/ST2/01074 of the National Science Center in Poland. He thanks the French LABEX P2IO and the French GDR QCD for support. 
\end{acknowledgments}


\end{document}